\documentclass[10pt]{article}

\def\centerbmp#1#2#3{\vskip#2\relax\centerline{\hbox to#1{\special
  {bmp:#3 x=#1, y=#2}\hfil}}}

\usepackage{latexsym}
\usepackage[latin2]{inputenc}
\usepackage{times}
\usepackage[normalem]{ulem}
\usepackage{color}

\newcommand{\qed}{$\Box$}
\newtheorem{theorem}{Theorem}

\newtheorem{corollary}[theorem]{Corollary}

\begin{document} 

\title{A Note on the PageRank of Undirected Graphs} 
\author{Vince Grolmusz\,$^{\rm a, b}$
\\
\small $^{\rm a}$ Protein Information Technology Group, Eötvös University,\\
 \small Pázmány Péter stny. 1/C, H-1117 Budapest, Hungary\\
\small $^{\rm b}$ Uratim Ltd.,  H-1118 Budapest, Hungary
}

\maketitle

\date{}

\begin{abstract}
The PageRank is a widely used scoring function of networks in general and of the World Wide Web graph in particular. The PageRank is defined for directed graphs, but in some special cases applications for undirected graphs occur. In the literature it is widely noted that the PageRank for undirected graphs are proportional to the degrees of the vertices of the graph. We prove that statement for a particular personalization vector in the definition of the PageRank, and we also show that in general, the PageRank of an undirected graph is not exactly proportional to the degree distribution of the graph: our main theorem gives an upper and a lower bound to the L$_1$ norm of the difference of the PageRank and the degree distribution vectors.
\end{abstract}

\paragraph*{Introduction} 

In this short note we are examining the PageRank \cite{Brin98theanatomy} of the undirected graphs. While the PageRank is usually applied for directed graphs (e.g., for the World Wide Web), in the literature it is sometimes mentioned in connection with undirected graphs \cite{Ivan2011}, \cite{Fortunato2}, \cite{Andersen}, \cite{WangLiuWang}, \cite{Abbassi}. 

In the literature it is frequently noted, that the PageRank of the vertices of an undirected graph is simply proportional to their degree. We intend to refine this statement here.

Let $G(V,E)$ denote an $n$-vertex undirected graph with vertex-set $V$ and edge-set $E$. Let $B=\{b_{ij}\}$ denote the $n\times n$ adjacency matrix of graph $G(V,E)$: that is, if the vertices are $V=\{v_1,v_2,\ldots,v_n\}$, then 
$$
b_{ij}=\cases{
1& if $v_i$ is connected to $v_j$, \cr
0 & \\ otherwise.\cr}
$$
Matrix $A$ is derived from matrix $B$ by dividing every entry of row $i$ by the degree of vertex $v_i$; therefore the sum of every row of $A$ equals to 1 ({\sl i.e.,} $A$ is a row-stochastic matrix). Note, that while matrix $B$ is symmetric, matrix $A$ is usually not. 

Suppose that $G(V,E)$ is a connected, undirected, non-bipartite graph. 

It is well known \cite{randomwalks} that the limit probability distribution of the random walk on the vertices of graph $G$, corresponding to the transition matrix $A$, exists and unique, and it is given by the column-vector 

$$f=\left({d(v_1)\over2|E|},{d(v_2)\over2|E|},\ldots,{d(v_n)\over2|E|}\right)^T$$

where $d(u)$ denotes the degree of vertex $u$. $f$ is called the degree-distribution vector of $G$. Clearly, 
$$A^Tf=f.$$
 
The PageRank \cite{Brin98theanatomy} of graph $G$ is defined as the limit probability distribution $\pi$ of the random walk, defined by the column-stochastic transition matrix  

$$cA^T+(1-c)v{\mathbf 1}^T,\eqno{(1)}$$

where $0<c<1$ is the damping constant, and vector $v$ with non-negative coordinates, satisfying ${\mathbf 1}^Tv=1$, is the personalization vector. 

The limit distribution $\pi$, called the PageRank vector, exists for all graph $G$, and it satisfies

$$(cA^T+(1-c)v{\mathbf 1}^T)\pi=\pi.\eqno{(2)}$$

Since ${\mathbf 1}^T\pi=1$ (2) can be written as

$$(I-cA^T)\pi=(1-c)v.$$

Matrix $I-cA^T$ is strictly diagonally dominant, therefore it is non-singular by the Gershgorin circle theorem \cite{Gershgorin, Golub, KamHav}:

$$\pi=(1-c)(I-cA^T)^{-1}v.\eqno{(3)}$$

With the eigenvector $f$ of $A$, satisfying $A^Tf=f$, we can write $(I-cA^T)f=(1-c)f$, or $f=(1-c)(I-cA^T)^{-1}f$, therefore if in the definition (1) of the PageRank, $v=f$, then 
$$\pi=f=(1-c)(I-cA^T)^{-1}f.\eqno{(4)}$$

So we proved the following 

\begin{corollary}
If the personalization vector $v$ in the definition of the PageRank of the undirected graph $G$ equals to the degree-distribution vector $f$ of $G$, then the PageRank vector $\pi$ equals to $f$, that is, the PageRank is exactly proportional to the degree distribution of the graph.\qed
\end{corollary}

We are interested in estimating the L$_1$-norm of the difference of vectors $f$ and $\pi$ for a general personalization vector $v$. From (3) and (4):
$$(\pi-f)=(1-c)(I-cA^T)^{-1}(v-f)\eqno{(5)}$$
and clearly
$$(I-cA^T)(\pi-f)=(1-c)(v-f).\eqno{(6)}$$

From \cite{KamHav}, 

$$\|I-cA^T\|_1=1+c, \ \ \|(I-cA^T)^{-1}\|_1={1\over1-c}.$$

Now, from (5):
$$\|(\pi-f)\|_1=(1-c)\|(I-cA^T)^{-1}(v-f)\|_1\leq \|v-f\|_1.\eqno{(7)}$$

Similarly, from (6):

$$(1-c)\|(v-f)\|_1= \|(I-cA^T)(\pi-f)\|_1\leq\|(\pi-f)\|_1(1+c).\eqno{(8)}$$

Inequalities (7) and (8) imply the following

\begin{theorem}
Let $A$ be a row-stochastic matrix, and let $f$, with non-negative coordinates, satisfy $A^Tf=f$, $\|f\|_1=1$. Let $\pi$ satisfy $(cA^T+(1-c)v{\mathbf 1}^T)\pi=\pi$ with $\|\pi\|_1=1$, $0<c<1$ and with $v$ of non-negative coordinates, satisfying ${\mathbf 1}^Tv=1$. Then
$${1-c\over1+c}\|v-f\|_1\leq \|\pi-f\|_1\leq \|v-f\|_1.$$\qed
\end{theorem}

Our theorem shows that the PageRank of an undirected graph equals to the degree-distribution vector if and only if $v=f$. In the original, non-personalized definition of the PageRank,
$$v={1\over n}{\mathbf 1},$$
therefore the original, non-personalized PageRank is equal to the degree-distribution vector $f$ of the undirected graph $G$ in case of regular graphs only.

\medskip

\noindent{\bf Acknowledgement.} The author acknowledges the partial support of the OTKA CNK 77780 and the TÁMOP 4.2.1./B-09/KMR-2010-0003 grants. The author is also grateful for an anonymous referee for his/her comments and suggestions.

\medskip

\noindent "Google" and "PageRank" are registered trademarks of Google, Inc.

\end{document}